\documentclass[lettersize, journal]{IEEEtran}
\usepackage{amsmath,amsfonts}
\usepackage{algorithm}
\usepackage{array}
\usepackage{textcomp}
\usepackage{stfloats}
\usepackage{url}
\usepackage{verbatim}
\usepackage{graphicx}
\usepackage{cite}
\usepackage{lipsum}
\usepackage{mathtools}

\usepackage{setspace}
\usepackage[utf8]{inputenc}
\usepackage{stackengine}
\usepackage{soul}
\usepackage{subfigure}
\usepackage[subfigure]{tocloft}
\usepackage{amsthm} 
\usepackage{float}
\usepackage{caption}

\usepackage{algpseudocode}

\begin{document}

\title{Optimization Theory Based Deep Reinforcement Learning for Resource Allocation in Ultra-Reliable Wireless Networked Control Systems}
\author{Hamida Qumber Ali, Amirhassan Babazadeh Darabi, Sinem Coleri, \IEEEmembership{Fellow,~IEEE}\vspace*{-\baselineskip}
\thanks{H. Q. Ali, A. B. Darabi and S. Coleri are with the Department of Electrical and
Electronics Engineering, Koc University, Istanbul, e-mail: {hirfan, adarabi22, scoleri}@
ku.edu.tr. Sinem Coleri acknowledges the support of the Scientific and Technological
Research Council of Turkey 2247-A National Leaders Research
Grant \#121C314.}}

\markboth{Submitted to IEEE Transactions on Communications}%
{Shell \MakeLowercase{\textit{et al.}}: A Sample Article Using IEEEtran.cls for IEEE Journals}

\IEEEpubid{0000--0000/00\$00.00~\copyright~2021 IEEE}

\maketitle

\begin{abstract}
    The design of Wireless Networked Control System (WNCS) requires addressing critical interactions between control and communication systems with minimal complexity and communication overhead while providing ultra-high reliability. This paper introduces a novel optimization theory based deep reinforcement learning (DRL) framework for the joint design of controller and communication systems. The objective of minimum power consumption is targeted while satisfying the schedulability and rate constraints of the communication system in the finite blocklength regime and stability constraint of the control system. Decision variables include the sampling period in the control system, and blocklength and packet error probability in the communication system. The proposed framework contains two stages: optimization theory and DRL. In the optimization theory stage, following the formulation of the joint optimization problem, optimality conditions are derived to find the mathematical relations between the optimal values of the decision variables. These relations allow the decomposition of the problem into multiple building blocks. In the DRL stage, the blocks that are simplified but not tractable are replaced by DRL. Via extensive simulations, the proposed optimization theory based DRL approach is demonstrated to outperform the optimization theory and pure DRL based approaches, with close to optimal performance and much lower complexity.  
\end{abstract}
	
   
\section{Introduction}
\IEEEPARstart{T}{he} recent advancements in the field of computing, sensing, control, and wireless technologies have drawn enormous interest in Wireless Networked Control Systems (WNCSs) that support emerging applications such as cyber-physical systems \cite{castro2022cyber}, Internet of Things \cite{bello2014intelligent}, and Tactile Internet \cite{promwongsa2020comprehensive}. WNCSs are spatially distributed control systems in which a physical plant is attached to sensor nodes that measure and transmit the plant state to the controller over a wireless channel. The controllers process the received information and forward their feedback to the actuators in order to control the physical plant \cite{hespanha2007survey}, \cite{wncs_survey}, \cite{al2016design}. WNCS plays a key role in Industry 4.0 \cite{Kager11} and several international organizations such as the Wireless Avionics Intra-Communications Alliance \cite{WinNT}, Zigbee Alliance \cite{ZigBee12}, International Society of Automation \cite{international14}, Highway Addressable Remote Transducer communication foundation \cite{WHart15}, and Industry IoT Consortium \cite{IIOT}. The main challenge in the design of wireless control systems is to optimize the performance of control and communication systems together while considering the critical interaction between the parameters of the two systems that directly influence the efficiency of the control systems and the lifetime, throughput and reliability of the wireless communication systems.
\IEEEpubidadjcol

Earlier research on the design and optimization of WNCS mostly focuses on model based methods, where mathematical models are used to model the performance of control and communication systems, and the solution methodologies employ either interactive design-based approach or joint design-based approach. The interactive design focuses on tuning the parameters of the wireless networks to meet the prespecified requirements of the control systems \cite{montalban2020noma}, \cite{dobslaw2014end}. The joint design of WNCS further considers the crucial interaction between the sampling period of the control system and the parameters of the wireless network such as the transmission power and rate at the physical layer, channel access mechanism at the MAC layer, and routing paths at the network layer in the formulation of the optimization problem. Due to the extensive complexity of the joint approach, earlier research was mostly focused on the usage of numerical techniques \cite{wu2010parameter}, \cite{pereira2007widom}. Later research has focused on providing the abstractions of control and communication systems to formulate the joint optimization problem [17]-[21]. In the joint optimization frameworks with the unique abstraction of the parameters of control and communication systems, the control system performance is formulated in terms of stochastic maximum allowed transfer interval (MATI) and the maximum allowed packet delay (MAD) constraint \cite{sadi2014minimum}, \cite{sadi2017joint}. Stochastic MATI is defined as guaranteeing the time interval between the reception time of subsequent state vector reports above MATI value with a prespecified probability whereas MAD ensures that the packet transmission delay does not exceed a predefined maximum limit. Other abstractions of control systems include Age-of-Information (AoI) \cite{zhao2022deep}, Age of Loop (AoL) \cite{de2021age} and Value of Information (VoI) \cite{farjam2021effect}.  None of these studies incorporate ultra-reliability requirement into the joint design of control and communication systems. However, in many emerging WNCS applications, sensors and other devices generate small packets carrying critical information that is expected to be received with low latency and ultra-high reliability, which requires incorporating finite blocklength information theory into the optimization framework. Moreover, the runtime complexity of the proposed algorithms may not be tolerated within the strict latency limitation of the control systems due to the usage of  optimization theory based iterative or heuristic solution methodologies. 

Recent model based techniques developed for the resource allocation in WNCS incorporates finite blocklength information theory into their optimization framework \cite{ren2019joint}-\cite{subchan2021energy}. \cite{ren2019joint} jointly optimizes the blocklength and power allocation for a factory automation scenario where a central controller transmits small packets to a robot and an actuator with the objective of minimizing the decoding error probability of the actuator subject to the reliability requirement of the robot as well as total energy and latency constraints. 
\cite{subchan2021energy} proposes an optimal estimator at the controller based on Kalman Filter and Linear Quadratic Regulator, and an optimal power scheduler to minimize the energy per symbol. These works do not consider the sampling period of the control system in the formulation of the joint optimization. Furthermore, the high complexity of the proposed model based methods may prevent their usage in low latency WNCS applications.

Due to the high complexity of model-based methods and the high overhead of collecting environmental parameters, machine learning-based approaches have recently been proposed for the joint design of control and communication systems \cite{lima2022model}, \cite{zhao2022deep}, \cite{baumann2018deep}. \cite{lima2022model} proposes a RL-based algorithm for Markovian control and resource allocation policies with the objective to minimize the deviations of the plant states from the equilibrium point. \cite{zhao2022deep} proposes a DRL-based algorithm for controller and scheduler optimization. A scheduler at the controller schedules only necessary transmissions from the sensor in order to reduce the uplink transmission rate while guaranteeing a required level of control quality. \cite{baumann2018deep} proposes two approaches  for learning resource-aware control policy by using a DRL-based algorithm; the first approach is based on an end-to-end learning of the structure of the communication and controller while the underlying policy outputs the communication decision and the control input. The second approach separately learns the communication strategy, and the stabilizing controller. These studies are completely data based without using any model of communication and control system, therefore, may require a large amount of training data in a dynamic environment.

In this paper, we propose an optimization theory based deep reinforcement learning algorithm for the resource allocation in wireless networked control systems, for the first time in literature. The resource allocation aims to determine the optimal values of all the parameters of the control and communication systems, including sampling period, blocklength, and packet error probability, with the goal of minimizing total power consumption. The constraints include the stochastic MATI and MAD constraints of the control systems, and the maximum transmit power and schedulability constraints of the communication system. The proposed framework utilizes the optimality conditions of mathematical system models in DRL to decrease the required amount of training data in DRL and enhance robustness against the non-idealities of the mathematical models. The novel contributions of this paper are presented below:
\begin {itemize}
\item We propose an optimization theory based DRL framework for the joint optimization of controller and communication systems by considering the ultra-reliable transmission in the finite blocklength regime, for the first time in literature. The framework includes the formulation of the optimization problem based on the abstractions of the communication and control systems while considering the ultra-reliable transmission in the finite blocklength regime, the derivation of the optimality conditions based on this formulation, and the replacement of the specific parts of the optimization problem that are not tractable by DRL.
\item We propose an optimization theory based DRL for the optimization of the sampling period, blocklength and packet error probability in the co-design of control and communication systems, for the first time in the literature. We perform this task in two stages, optimization theory stage and DRL stage. In the optimization theory stage, we first derive the optimality conditions among the sampling period, blocklength and packet error probability by using the model based approach. This allows us to reduce the decision variables to the blocklength only. In the DRL stage, we use DNN to formulate the remaining simplified problem. 
\item We analyze the performance of the proposed optimization theory based DRL approach in comparison to the optimization theory model based approach and pure DRL approach via extensive simulations, for different network sizes, environments, stochastic MATI and MAD constraints.
\end {itemize}
This paper is structured as follows: Section \ref{systemmodel} presents control and communication system models and the underlying assumptions used in this paper. Section \ref{jointoptimization} presents the formulation of the joint optimization of controller and communication systems. Section \ref{optimizationtheory_drl} introduces the proposed optimization theory based DRL approach. Section \ref{performanceevaluation} presents the performance evaluation of the proposed approach in comparison to the optimization theory and pure DRL based benchmark algorithms. Section \ref{conclusion} concludes the paper. 

\section {System Model and Assumptions} \label{systemmodel}
We assume a WNCS architecture comprising of multiple plants that are constantly monitored by a central controller over an unreliable wireless network. Multiple sensors attached to each plant are responsible of periodically measuring the plant state. We assume that the size of the packets carrying plant state information is extremely low, usually smaller than 20 bytes, to support the low-latency transmission \cite{yilmaz2015analysis}. Moreover, the transmitted data may not be received by the controller due to the unreliable wireless communication channel. Based on the most recent received state update information, the controller computes a new control command and transmits it to the actuator. We assume that the actuator is collocated with the controller due to the criticality of the control command \cite{arampatzis2005survey} and therefore, the control command is always received successfully.

The sensors monitoring a plant send periodic state updates to the controller. The sampling period, the blocklength, the packet transmission delay and the packet error probability of node $i$ are denoted by $h_i$, $m_i$, $d_i$ and $p_i$, respectively, for $i \in \{1,2,...,N\}$. We assume that $d_i  \leq h_i $ and the packets arriving later than the sampling period are considered as outdated and discarded in order to avoid out of order arrival problem at the controller. The outdated packets are not retransmitted because outdated state information is useless or even harmful for time-critical control applications. The packet error is modeled as a Bernoulli random process to simplify the problem.

We assume Time Division Multiple Access (TDMA) as a channel access mechanism since TDMA ensures a deterministic access delay and is widely preferred for many industrial control applications \cite{ANSI38}. We assume that the channel time is divided into frames. Each frame is further divided into time slots in which the first slot is reserved for beacon frame. The controller transmits the beacon periodically to broadcast the synchronization and scheduling updates among the nodes within the WNCS. In the scheduling update, nodes are assigned time slots for their corresponding data transmissions along with other parameters including the optimal transmission power, blocklength, and sampling period. We assume that nodes within the network do not transmit concurrently and the packet error rate is continually monitored by the network manager. 
We assume that the radio of each node can operate in three different modes; active mode in which it can sense the channel, transmit and receive a packet; sleep mode in which significant parts of the transceiver are switched off and the transient mode in which the node transitions from sleep to active state or vice versa. When the nodes are not scheduled to transmit or receive a packet, they switch to the sleep mode. We only consider the energy consumption in the transmission of packets in the active mode because the sleep and transient modes consume very small energy compared to active mode \cite{cui2005energy} and the beacon packets consume fixed energy.

Next, we discuss the system models for the control and communications systems based on the requirements and constraints related to the stability of the control system and the performance of the wireless communication system.
\vspace*{-\baselineskip}
\subsection{Control System Model}
We formulate the stability and performance conditions of the control system in terms of stochastic $MATI$ and $MAD$ as explained in detail in \cite{sadi2017joint}, and widely used in many control applications, such as wireless industrial automation \cite{ANSI38}, air transportation systems \cite{park2014high}, and autonomous vehicular systems \cite{karagiannis2011vehicular}.
\subsubsection{Stochastic MATI Constraint}
The stochastic MATI constraint is defined as the maximum allowed time interval between the reception of subsequent state vector reports above MATI value with a predefined probability. This constraint is formulated as 
\begin{equation}
    P[ \mu_i(h_i, d_i(m_i), p_i) \leq \Omega] \geq \delta.
\end{equation}
where $\mu_i$ denotes the time interval between two subsequent state vector updates of sensor node $i$, $\Omega$ denotes the stochastic MATI and $\delta$ indicates the minimum probability with which MATI should be achieved. $\mu_i$ is a function of the sampling period $h_i$, packet transmission delay $d_i$ and packet error probability $p_i$. The delay $d_i$ is a function of the blocklength $m_i$ which denotes the number symbols used to transmit packet of length $L_i$ bits. In order to satisfy $\mu_i \leq \Omega$, there must be at least one successful transmission within $\Omega$ duration. 
 For given values of $h_i$ and $\Omega$, there are $\lfloor \frac{\Omega}{h_i} \rfloor$ opportunities of the state vector update reception. We model the packet error $p_i$ as a Bernoulli random process and rewrite the stochastic MATI constraint given in Eq. (1) as
\begin{equation}
1 - p_i^{\lfloor \frac{\Omega}{h_i} \rfloor} \geq \delta  \label{seq2}. \\
\end{equation}
The values of $\Omega$ and  $\delta$ are determined by the underlying applications.
\subsubsection{MAD Constraint}
The MAD constraint is defined as the maximum packet transmission delay not exceeding a predefined maximum limit. This constraint is included in order to stabilize the control system and ensure its efficiency. MAD is formulated as
\begin{equation}
    d_i(m_i) \leq \Delta,
\end{equation}
where $\Delta$ denotes MAD. The underlying control system determines the value of MAD.
\vspace*{-\baselineskip}
\subsection{Communication System Model}
We formulate the blocklength, power consumption, transmit power, and packet error probability constraints of the underlying communication system in the finite blocklength regime. 
\subsubsection{Blocklength Constraint}
The plant state information measured by node $i$ is transmitted in the form of small packets of length $L_i$ by using blocklength $m_i$ such that 
\begin{equation}
 m_i \leq M_{th}, 
\end{equation}
where $M_{th}$ is the maximum allowed number of channel uses or maximum delay, within which the packet transmission has to finish. The coding rate is defined as the ratio of the number of information bits to the number of symbols used to transmit them. The Shannon capacity is defined as the highest coding rate of a communication system, provided that an encoder/decoder pair exists whose decoding error probability becomes negligible when the blocklength approaches infinity \cite{shanon}. However, in URLLC, the blocklength for each frame is small and the decoding error probability cannot be ignored. The coding rate $R_i$ (in bits/sec/Hz) of node $i$ under finite blocklength regime can be approximated as \cite{polyanskiy2010channel}
\begin{equation}
   R_i \approx \log_2(1+\gamma_i) - \sqrt{\frac{V_i}{m_i}} \frac{Q^{-1}(\epsilon_i)}{ln{(2)}},
\end{equation}
where $\gamma_i= \frac{W_{tx,i} \lvert g_i \rvert }{\sigma^2}$ is the signal-to-noise ratio (SNR) at the controller; $W_{tx,i}$ is the transmit power of node $i$; $g_i$ is the channel gain from the node $i$ to the controller; $\sigma^2$ is the noise power; $V_i=1-(1+\gamma_i)^{-2}$ is the channel dispersion; $\epsilon_i$ is the decoding error probability and $Q^{-1}$ denotes the inverse of the Gaussian $Q$ function. In short frame structure, the frame duration is smaller than the channel coherence time, called a quasi-static channel, which means that the symbols in a packet experience similar fading states. Decoding an information bit wrongly is, therefore, considered as the only factor making a packet erroneous. For this reason, the decoding error probability has been interchangeably used for packet error probability \cite{durisi2016toward}, \cite{ren2019joint}. We can safely replace $\epsilon_i$ by $p_i$ in our equations in the rest of this paper.
\subsubsection{Power Consumption}
Average power consumption $W_i$ of node $i$ is a function of the sampling period $h_i$, packet error probability $p_i$, and delay $d_i(m_i)$ such that
\begin{equation}
    W_i(h_i,d_i(m_i),p_i) = \frac{(W_{tx,i} + W_i^c)d_i(m_i)}{h_i},
\end{equation}
where 
\begin{equation}
    d_i(m_i) = \frac{m_i}{B},
\end{equation}
$W_i^c$ is the circuit power consumption when the transmitter is in the active mode and B is the bandwidth of the wireless channel. We derive the packet error probability $p_i$ of node $i$ from Eqn. (5) as 
\begin{equation}
    p_i = Q\left[ \ln{(2)} \sqrt{\frac{m_i}{V_i}} \left(log_2(1 + \frac{W_{tx,i} \lvert g_i \rvert }{\sigma^2}) - \frac{L_i}{m_i} \right) \right].
\end{equation}
The transmit power $W_{tx,i}$ of node $i$ can be derived from Eqn. (8) as
\begin{equation}
  W_{tx,i} = \frac{\sigma^2}{|g_i|} \left[\exp\left( \frac{Q^{-1} (p_i)}{\sqrt{m_i}} + \frac{\ln(2)L_i}{m_i}\right) - 1 \right].  
\end{equation}
Now Eqn. (6) can be rewritten as
\begin{align}
W_i(h_i, d_i(m_i), p_i) &= \frac{m_i \sigma^2}{h_i B |g_i|} \nonumber \\
&\quad \left[\exp\left( \frac{Q^{-1} (p_i)}{\sqrt{m_i}}  +  \frac{\ln(2)L_i}{m_i}  \right) - 1\right] \nonumber \\
&\quad + \frac{W_i^c m_i}{h_i B},
\end{align}
where the channel dispersion is approximated by $V_i \approx 1, \forall i \in \mathbb {N}$, which is widely used in ultra-reliable communication for medium-to-high values of SNR to make the problem tractable \cite{ren2019joint}. 
\subsubsection{Maximum Transmit Power Constraint}
Each node can use a maximum power level, denoted by $W_{tx,max}$, for its transmissions. We formulate the maximum transmit power constraint as
\begin{equation}
     W_{tx,i} \leq  W_{tx,max}.
\end{equation}
\subsubsection{Schedulability Constraint}
The schedulability constraint handles the assignment of transmission times to multiple sensor nodes in a network considering wireless transmission constraints. We assume that no two nodes can transmit concurrently and generate transmission schedule by the Earliest Deadline First (EDF) algorithm which is a dynamic scheduling algorithm prioritizing transmissions closest to their deadlines \cite{Dertouzos46}.
When the maximum allowed packet transmission delay and the packet generation period are not equal, i.e. $ \exists  i \in [1,N], \Delta \neq h_i$, the simulation of the EDF algorithm can generate an exact schedule based on $d_i, h_i, \Delta$ and task arrival times $a_i$ for all $i \in [1,N]$ which is feasible if and only if no deadlines are missed \cite{Zhang50}, \cite{eisenbrand2010edf}. For an optimization framework, however, we need to explicitly formulate the necessary and sufficient conditions for schedulability. We define the schedulability constraint as proposed by \cite{sadi2014minimum}
\begin{equation}
     \sum_{i=1}^{N} \frac{d_i(m_i)}{h_i} \leq \beta, 
\end{equation}
where $\beta$ is defined as the utilization bound which should satisfy $ 0 < \beta \leq 1$. Eqn. (12) is a necessary and sufficient condition for a feasible schedule for $\beta =\beta_{nec}=1$ and $\beta=\beta_{suf}=min(1,min_{i \in [1,N]} \frac{\Delta}{h_i})$, respectively \cite{Zhang50}. Each node $i$ in the network is allocated a ratio of the total schedule length $\frac{d_i}{h_i}$ for its transmission. We assume that no two nodes can transmit concurrently, so the sum of these terms equals to the ratio of total time assigned to all the nodes $i\in [1, N]$ in the network to the schedule length. $\beta$ can not exceed one because the total time allocated to all the nodes in the network should be within the schedule length. 

\section{Joint Optimization of Control and Communication Systems} \label{jointoptimization}
\singlespacing In this section, we formulate the problem of the joint optimization of control and communication systems for ultra-reliable communication in the finite blocklength regime with the objective of minimizing the power consumption of the communication system. The model considers the stochastic MATI and MAD constraints to guarantee the stability of the control system, and maximum transmit power, maximum blocklength, power consumption and schedulability constraints of the wireless communication system. The joint optimization of control and communication systems is formulated as follows:
\begin{subequations}
\begin{align}
\underset{\substack{h_i, m_i, p_i \\ i \in [1,N]}}{\text{min}} \quad &\sum_{i=1}^{N} C_{i1} C_{2} \frac{m_i}{h_i}\nonumber\\
&\left[\exp\left( \frac{Q^{-1} (p_i)}{\sqrt{m_i}} + \frac{\ln(2)L_i}{m_i}\right) - 1\right]\label{eq:first}\\
& + \frac{C_{2} W_i^c m_i }{h_i}\label{eq:second}
\end{align}
\text{s.t.}
 \begin{equation}
{\lfloor \frac{\Omega}{h_i} \rfloor} \ln{p_i} - \ln(1-\delta)
\leq 0 , \forall i \in [1,N]
\end{equation}
\begin{equation}
 0 < d_i(m_i) \leq min(\Delta, h_i), \forall i \in [1,N] \\ 
\end{equation}
\begin{equation}
 0 < h_i \leq \Omega, \forall i \in [1,N]  \\ 
\end{equation}
\begin{equation}
  0 < p_i < 1, \forall i \in [1,N] \\ 
\end{equation}
\begin{equation}
 \quad  m_i \leq M_{th}, \forall i \in [1,N] 
\end{equation}
\begin{equation}
  C_{i1} \left[\exp\left( \frac{Q^{-1} (p_i)}{\sqrt{m_i}} + \frac{\ln(2)L_i}{m_i}\right) - 1\right] \leq W_{tx,max} 
\end{equation}
\begin{equation}
    \sum_{i=1}^{N} \frac{d_i(m_i)}{h_i} \leq \beta \label{seq6}, 
\end{equation}
  
\end{subequations} 
where the variables $C_{i1} = \frac{\sigma^2}{|g_i|}$ and $C_{2} = \frac{1}{B}$. The decision variables of the optimization problem are the sampling period $h_i$, packet error probability $p_i$ and blocklength $m_i, i \in [1,N]$. Eqn. (13a) gives the objective of the problem for the minimization of the total power consumption, as derived in Eqn. (10). Eqn. (13b) represents the stochastic MATI constraint. Equation (13c) combines the MAD constraint with the requirement that the packets arriving later than the sampling period are considered as outdated, i.e. $d_i \leq h_i$. Eqn. (13d) states that the sampling period $h_i, i\in [1,N]$ must be less than MATI. Equations (13e), (13f), (13g), and (13h) represent the threshold constraint of the lower and upper bounds for the packet error probability, blocklength limit, the maximum transmit power constraint and the schedulability constraint, respectively.

This optimization problem is a non-convex Mixed-Integer Programming problem, thus, any solution for global optimum is difficult \cite{Convex47}.

\section {Optimization Theory Based Deep Reinforcement Learning} \label{optimizationtheory_drl}
The optimization theory based deep reinforcement learning is based on the hybrid usage of the optimization theory based solution methodology using domain knowledge and deep learning techniques that learn from data.  The approach consists of two stages: 1) Optimization theory stage: Following the formulation of the optimization problem based on the domain knowledge from the system models of the control and communication systems, the solution is decomposed into multiple building blocks based on the derivation of the optimality conditions. 2) DRL stage: The building blocks that are complicated or intractable are replaced by deep learning structure, which is then trained based on data. This hybrid usage of model and data based approaches allows us to reduce the amount of training data used in model agnostic DRL. Next, these optimization and DRL stages are provided in Sections \ref{opt_theory_stage} and \ref{DRL_stage}, respectively.
\vspace*{-\baselineskip}
\subsection{Optimization Theory Stage}\label{opt_theory_stage}
In the optimization theory stage, the general solution is decomposed into multiple building blocks. The optimization theory enables the derivation of the optimality conditions on the sampling period, packet error probability and the blocklength. This allows the reformulation of the problem in terms of only one variable, blocklength, as an Integer Programming (IP) problem. The remaining variables, i.e. sampling period and packet error probability, can then be obtained by using the optimality conditions. 
Next, we formulate the optimality conditions to find the relation between the optimal sampling period and the optimal packet error probability.

 \textbf{Lemma 1:} Let us denote the optimal values of the sampling period, packet error probability and blocklength by $h_i^*$, $p_i^*$ and $m_i^*$, respectively. The relation between the optimal values of the sampling period and the packet error probability is given by
\begin{equation}
   \frac{\Omega}{h_i^*} = \frac{\ln(1-\delta)}{\ln{p_i^*}} = k_i,
   \end{equation}
where $k_i$ is a positive integer. 

\vspace{-.8em}
\begin{proof} We prove the Lemma by contradiction in a similar way as in \cite{sadi2014minimum}. We start by assuming that $\frac{\Omega}{h_i^*}$ is not a positive integer such that $\lfloor \frac{\Omega}{h_i^*} \rfloor < \frac{\Omega}{h_i^*}$. If we increase $h_i^*$ such that $\lfloor \frac{\Omega}{h_i^*} \rfloor = \frac{\Omega}{h_i^*}$ while the upper bound given in Eqn. (13d) is not violated, the constraint in Eqn. (13b) still holds since the value of $\lfloor \frac{\Omega}{h_i^*} \rfloor$ remains unchanged and the constraints (13c) and (13h) also hold with the increase in the value of $h_i^*$. However, the objective cost function given in Eqn. (13a) decreases since it is a monotonically decreasing function of $h_i$.
Similarly, $\frac{\Omega}{h_i^*} = \frac{\ln(1-\delta)}{\ln{p_i^*}}$ can be proved by contradiction. We assume that $\frac{\Omega}{h_i^*} < \frac{\ln(1-\delta)}{\ln{p_i^*}}$ and increase $p_i^*$ such that $\frac{\Omega}{h_i^*} = \frac{\ln(1-\delta)}{\ln{p_i^*}}$, the constraint in Eqn. (13g) still holds since the power consumption of node $i$ is non-increasing  function of $p_i$. However, the power consumption function in Eqn. (13a) decreases since it is a monotonically decreasing function of $p_i$.
\end{proof} 
\vspace{-1em}
Next, we express the optimal value of $k_i$ in terms of $m_i$. This enables us to express the optimization problem (13) in terms of only one variable $m_i$. 

\textbf{Lemma 2:} The optimal value of $k_i$ in Lemma 1, expressed by $k_i^*$, is as a function of $m_i$ and given by
\begin{equation}
\begin{split}
   k_i^*(m_i) = & max \\
   &\left[1,\frac{\ln(1-\delta)} {\ln \left[Q \left [\sqrt{m_i} \ln \left(\frac{W_{tx,max} }{m_i C_{i1}}+1 \right) - \frac{\ln{(2)}L}{\sqrt{m_i}} \right]  \right] } \right]. 
\end{split}
\end{equation}
\begin{proof}
The power consumption in Eqn. (13a) is a monotonically increasing function of $k_i$. Therefore, $k_i^*$ is the minimum positive integer that satisfies the constraints (13b)-(13e) and (13g). In Eqn. (15), the second term of the maximum function is the minimum value of $k_i$ that satisfies the constraint (13g) since constraint (13c) is satisfied by the minimum value of $k_i$ in case a feasible solution exists. The constraints (13b) and (13d)-(13f) follow as the constraints (13c) and (13g) are satisfied. 
\end{proof}

 \begin{figure}[!t]
     \centering
     \subfigure[Original Problem] {
         \includegraphics[width=.3\textwidth]{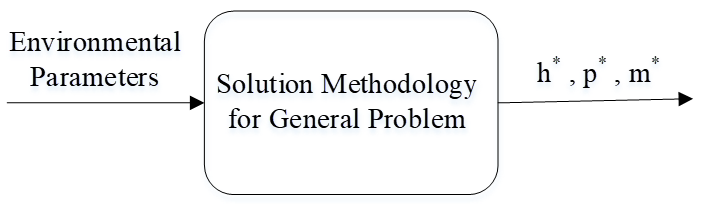}
          }
         \hspace*{-0.1em}
         \subfigure [Decomposed Problem] {
         \includegraphics[width=.45\textwidth]{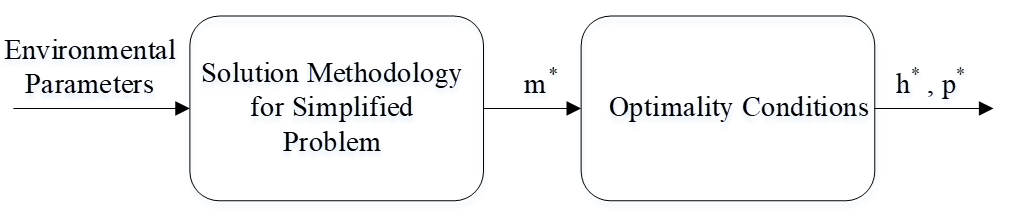}
          }
          \hspace*{-0.1em}
         \subfigure [Decomposed DRL based Problem] {
         \includegraphics[width=.43\textwidth]{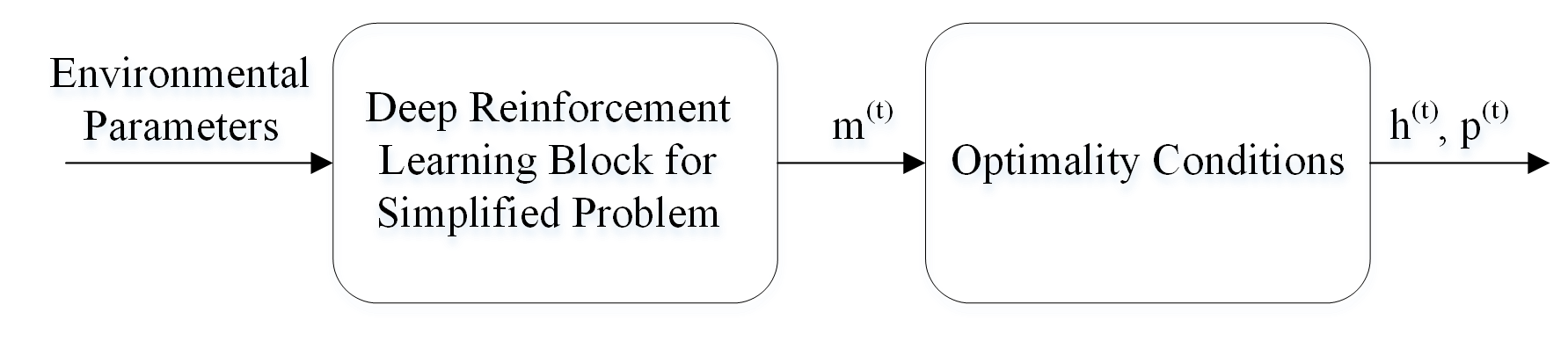}
          }
        \caption{Problem decomposition by optimality conditions}
        \label{fig:subfigures}
\end{figure}
By using (7) and (14), Eqn. (12) can be rewritten as 
  \begin{equation}
     \sum_{i=1}^{N} \frac{C_2 m_i  k_i^*(m_i)}{\Omega} \leq \beta. 
\end{equation}

The joint optimization problem (13) can then be reformulated as
\begin{subequations} 
\begin{align}
    &\underset{m_i, i \in [1,N]}{\text{min}} \sum_{i=1}^{N} C_{i1} C_{2} \frac{m_i k_i^*(m_i)}{\Omega} \nonumber \\
    &\quad \left[\exp\left(\frac{Q^{-1}(1-\delta)^{\frac{1}{k_i^*(m_i)}}}{\sqrt{m_i}} + \frac{\ln(2)L_i}{m_i}\right) - 1\right] \label{eq:first2} \\
    &\quad + \frac{C_{2} W_i^c m_i k_i^*(m_i)}{\Omega} \label{eq:second2}
\end{align}
\quad \text{s.t.} 
\begin{equation}
     m_i \leq M_{th}, \forall i \in [1,N]
\end{equation}
\begin{equation}
     \sum_{i=1}^{N} \frac{C_2 m_i k_i^*(m_i)}{\Omega} \leq \beta,
\end{equation}
\end{subequations}

 Fig. \ref{fig:subfigures} shows the decomposition of the optimization problem (17) into multiple building blocks based on the derivation of the optimality conditions. Let $m^*$, $h^*$ and $p^*$ denote the vectors containing the optimal values of the blocklength, sampling period and packet error probability given by $m_i^*$, $h_i^*$ and $p_i^*$ in the $i$th element, respectively. Fig. \ref{fig:subfigures}.a shows the solution methodology for the general problem, where environmental parameters are the input and the optimal values of blocklength, sampling period and packet error probability are the output. Fig. \ref{fig:subfigures}.b shows the decomposed problem, where in the first building block, the solution methodology is used for a simpler problem with environmental variables at the input and only optimal values of the block length at the output, and in the second building block, the optimal values of the blocklength are used to obtain the optimal values of the sampling period and packet error probability by using optimality conditions.
\subsection{DRL Stage}\label{DRL_stage}
In the DRL stage, the building blocks that are complicated or intractable are replaced by the deep learning block. Let $m^{(t)}$, $h^{(t)}$ and $p^{(t)}$ denote the vectors containing the values of the blocklength, sampling period and packet error probability for node $i$, given by $m_i^{(t)}$, $h_i^{(t)}$ and $p_i^{(t)}$, respectively, in the $i$th element at time step $t$. In Fig. 1.c, the deep learning based building block is used to obtain the optimal values of the blocklength in the optimization problem (17). The main motivation for using data driven deep learning based solution is handling the high complexity of the problem and providing robustness based on the real data.

Next, we provide the overview of deep reinforcement learning, the proposed DRL-based solution and the proposed DRL-based framework.
\subsubsection{Overview of Deep Reinforcement Learning}\label{DRL_overview}
A reinforcement learning agent aims to act in an optimal manner by having several interactions with its environment and getting rewards \cite{sutton2018reinforcement}. The tuple $s^{(t)} \in S$ is a state consisting of variables that have an impact on the changes in the environment at discrete time step $t$. The agent takes action $a^{(t)} \in A$ at time step $t$, following the policy, which can be either stochastic or deterministic, denoted by $a^{(t)} \sim \pi(.|s^{(t)})$ or $a^{(t)} = \mu(s^{(t)})$, respectively. For the stochastic case, the policy function should meet the condition $\sum_{a^{(t)} \in A} \pi(a^{(t)}|s^{(t)}) = 1$, where $A$ is the action space. After taking the action $a^{(t)}$, the agent moves from state $s^{(t)}$ to $s^{(t+1)}$ while getting a reward $r^{(t)}$. An experience is gathered at time $t$ that is a tuple in the form of $e^{(t)} = (s^{(t)}, a^{(t)}, r^{(t)}, s^{(t+1)})$.
Model-free reinforcement learning learns from interacting with the environment without any further information on the transition probabilities. 

Q-learning algorithm finds a policy to maximize the cumulative future reward. The Q-function is the expected future reward when an action is taken under the policy $\pi$, which is written as
\begin{equation} \label{eq:2}
    Q^{\pi}(s, a) = E_{\pi}[R^{(t)} | s^{(t)} = s, a^{(t)} = a],
\end{equation}
where $R^{(t)}$ is the future discounted reward defined as 
\begin{equation} \label{eq:1}
    R^{(t)} = \sum_{\tau=0}^{\infty}\gamma^{\tau}r^{(t + \tau)},
\end{equation}
where $\gamma \in (0, 1]$ is the discount factor. 
The conventional Q-learning algorithm builds a lookup table of Q-values. Once the Q-values are randomly initialized, the agent chooses actions based on $\epsilon$-greedy policy for each time step $t$. The $\epsilon$-greedy policy implies that the agent takes action $a'$, which maximizes the expected reward, with a probability of $1-\epsilon$, and exploits the environment. On the other hand, it explores the environment by taking a random action with the probability $\epsilon$. A decaying $\epsilon$-greedy algorithm, in which the $\epsilon$ value decreases after each iteration, is used to increase the number of exploitations over time. 

The complexity of the conventional Q-learning method increases as the size of the state space increases. A Deep Q-network (DQN) overcomes this problem by employing a deep neural network (DNN) to represent Q-function, which is expressed as $Q(s, a; \theta)$, in which $\theta$ is the network parameter. Like conventional Q-learning, a DQN collects experience by taking an action in a specific state. An experience replay buffer $\mathcal{D}$ stores different experiences in the memory. A mini-batch $\mathcal{E}^{(t)} = \{e_1^{(t)}, e_2^{(t)}, ..., e_{\mathcal{B}}^{(t)}\}$ is sampled from the experience replay buffer with length $\mathcal{B}$, where $e_i^{(t)}$ is the $i$-th experience sampled from the experience replay buffer. Sampling is performed in order to remove correlations in the observations and flatten changes in the data distribution. Furthermore, two DQNs are defined by using the ``quasi-static target network" method: The target network with parameters $\theta^{(t)}_{{target}}$ and the train network with parameters $\theta^{(t)}_{{train}}$. $\theta^{(t)}_{{target}}$ is updated with a soft update method every time frame such that
\begin{equation} \label{soft_update}
    \theta^{(t)}_{target} = \tau \theta^{(t)}_{train} + (1 - \tau)\theta^{(t)}_{target},
\end{equation}
where $\tau \ll 1$. The soft update method offers more stability and excludes extreme updates that differ from past experiences. The least squares loss of the train DQN for a randomly selected mini-batch $\mathcal{E}^{(t)}$ at time $t$ from the experience replay buffer is
\begin{equation} \label{eq:DQN_loss}
    \mathcal{L}(\theta^{(t)}_{{train}}) = \sum_{(s, a, r, s') \in \mathcal{E}^{(t)}} (y^{(t)}_{\textsc{DQN}}(r, s') - Q(s, a; \theta^{(t)}_{{train}}))^2
\end{equation}
where the target is
\begin{equation} \label{eq:7}
    y^{(t)}_{\textsc{DQN}}(r, s') = r + \lambda \max_{a'} Q(s', a'; \theta^{(t)}_{{target}}).
\end{equation}
The agent updates network parameters $\theta^{(t)}_{train}$ by using an optimization technique to minimize the expected prediction error (cost function) of the sampled mini-batch. The stochastic gradient descent utilizes the computed gradients and has been proved to converge to a set of suitable parameters fast. The gradient update for the network is performed as follows:
\begin{equation} \label{gradient_update}
    \theta_{train}^{(t)} \leftarrow \theta_{train}^{(t)} - \eta^{(t)} \nabla_{\theta_{train}^{(t)}} \mathcal{L}(\theta_{train}^{(t)}),
\end{equation} \\
where $\eta^{(t)} \in (0, 1)$ is the learning rate of the gradient update.

\textit{ \emph Double Q-Learning:} Conventional Q-learning suffers from an overestimation bias because of the maximization step in (\ref{eq:7}). Double Q-learning addresses this problem by decoupling, in the maximization performed for the bootstrap target, the selection of the action from its evaluation \cite{hasselt2010double}. 

\textit{ \emph Dueling networks:} The dueling network represents a neural network structure specifically created to facilitate value-based reinforcement learning \cite{hessel2018rainbow}. The dueling network incorporates two separate modules, namely the value and advantage estimators. These estimators are combined using a unique aggregator mechanism.

\begin{figure*}[!htb]
    \centering
    \includegraphics[width= 120mm, scale= 1.8]{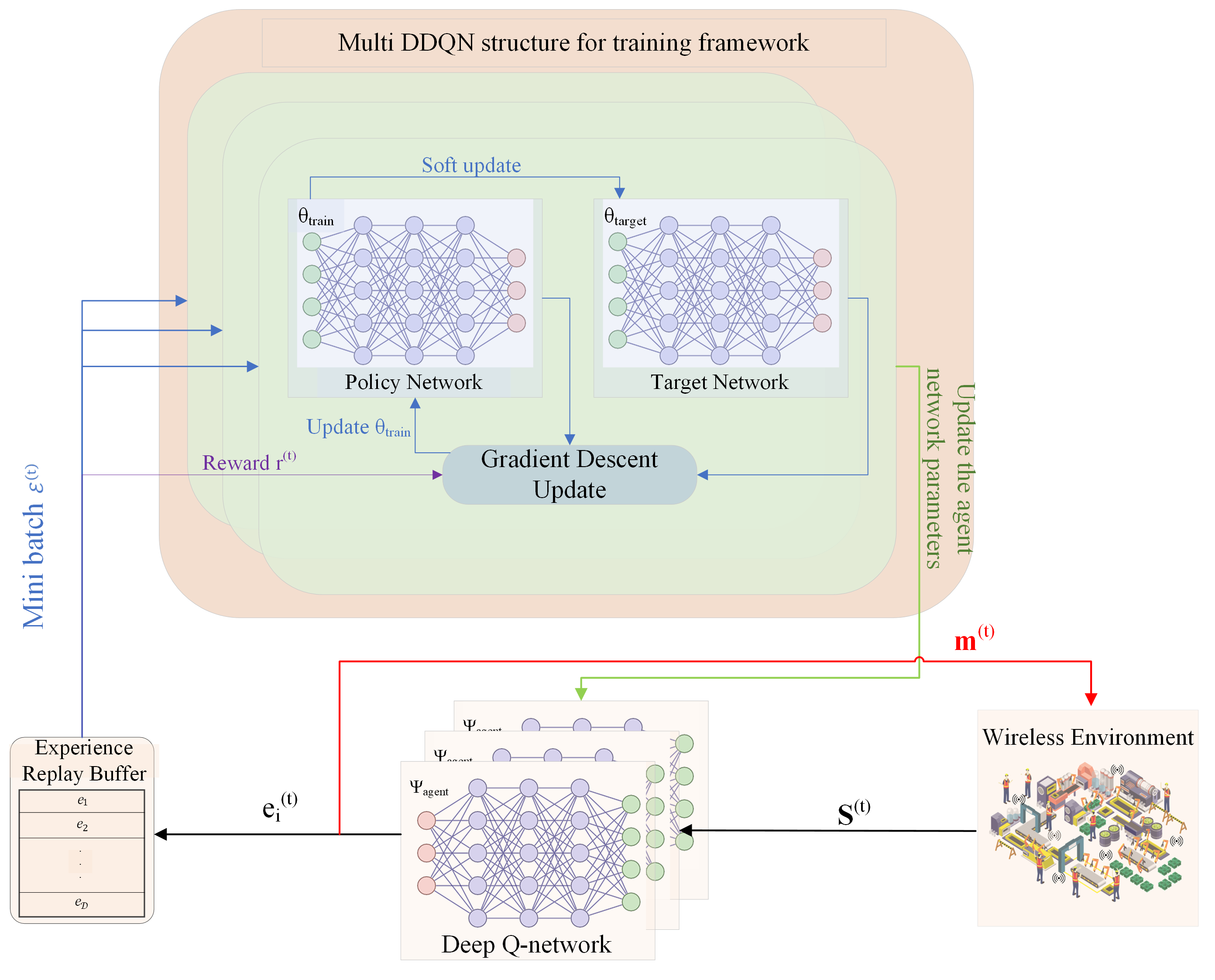}
    \caption{Proposed DRL algorithm.}
    \label{fig:arch}
\end{figure*}

\subsubsection{Proposed DRL Algorithm}\label{DRL_proposed_alg}
A centrally-trained-centrally-executed multi-agent deep reinforcement learning model is proposed in Fig. \ref{fig:arch}, with each sensor node having its local network in the central server as an agent. The changes in the states of the environment in the multi-agent learning system are dependent on the joint actions of the agents. All the agents work in a synchronized way and take their actions simultaneously. A centralized training and centralized execution method is used to smooth the implementation. Each agent $i$ observes the state $s_i^{(t)}$ and chooses an action $a_i^{(t)}$ using a local neural network and gets the reward $r_i^{(t)}$. Then, the experience of the agents is collected into a replay buffer, which works in a FIFO (first in, first out) fashion. Training is done based on the batches sampled from the buffer to update the network parameters. DQN and dueling double deep Q-network (DDQN) are implemented as the learning scheme for the agents. The elements of the implementation are described below.

\begin{itemize}
\item \textit{States:} States are the variables that directly have an impact on the environment. The next state of each agent is determined by the joint actions of all agents. The state of agent $i$ is given by
\begin{equation} \label{eq:8}
    s_i^{(t)} = (m^{(t-1)}, R_i^{(t-1)}, W^{(t-1)}, \mathcal{P}^{(t-1)}, \gamma_i^{(t)}, g^{(t)}).
\end{equation}
where $m^{(t-1)}$ is the blocklength vector in the previous time step, where the $i$th element of the vector corresponds to the chosen blocklength of the $i$th node, $m_i^{(t-1)}$; $R_i^{(t-1)}$ is the rate of node $i$ in the previous time step; $W^{(t-1)}$ is the power value vector in the previous time step, where the power of node $i$ is the $i$th element of the power value vector, $W_{tx, i}^{(t-1)}$; $\mathcal{P}^{(t-1)}$ is the total power consumption of the nodes; $\gamma_i^{(t)}$ is the SNR calculated based on the instantaneous channel gains and the power level in the previous time step, i.e., $\gamma_i^{(t)} = \frac{g_{i}^{(t)} W_{tx, i}^{(t-1)}}{\sigma^2}$; and $g^{(t)}$ is the instantaneous channel gain vector, where the $i$th element corresponds to the channel gain of node $i$, $g_{i}^{(t)}$. The parameters from the previous time step are used to form a Markov Decision Process (MDP) and constrain the model further for the blocklength. Rate and SNR are included to get as much information as possible about the channel condition.

    \item \textit{Action:} The agent $i$ determines the blocklength $m_i^{(t)}$ to send. All agents have the same action space given by
\begin{equation} \label{7_2}
    A = \{1, 2, ..., M_{th}\},
\end{equation}
which is determined from the optimization problem constraint related to the maximum blocklength available for an agent.

    \item \textit{Reward:} The reward function is the central core of the learning process in RL. Each agent takes optimal action to maximize the total reward. The reward function has two parts. The first part corresponds to the objective function of the optimization problem (17a) and is given by
\begin{align}\label{eq:9}
    r_g^{(t)} &= -\sum_{i=1}^{N} \left(C_{i1}^{(t)}C_{2}\frac{m_i^{(t)}k_i^{*(t)}}{\Omega}\left[\exp\left(\frac{Q^{-1}(1-\delta)^{\frac{1}{k_i^{*(t)}}}}{\sqrt{m_i^{(t)}}} \right.\right. \right. \nonumber \\
    &\quad + \left. \left. \frac{\ln{(2)}L_i}{m_i^{(t)}}\right) - 1\right] \\
    &\quad + C_2 W_i^c \frac{m_i^{(t)}k_i^{*(t)}}{\Omega}, \nonumber
\end{align}
where $C_{i1}^{(t)} = \frac{\sigma^2}{|g_i^{(t)}|}$. The second part of the reward function corresponds to the schedulability constraint (17c). If the agents violate the schedulability constraint, they will get penalized based on their distance to the satisfaction of this constraint. On the other hand, a small positive reward is gained if the agents satisfy the constraint. The second part is then written as

\begin{equation}
    r_{i, l}^{(t)} = 
    \begin{dcases}
        r_{pos}, & \quad \sum_{j=1}^N \frac{C_{2j}m_j^{(t)} k_j(m_j^{(t)})}{\Omega} \leq \beta \\
        r_{neg}, & \quad \text{otherwise,}
    \end{dcases}
\end{equation}
where $r_{pos} \in \mathbb{R}^+$ is a small positive constant reward for not violating the constraint, and $r_{neg} = \mathcal{C}(\sum_{j=1}^N \frac{C_{2j}m_j^{(t)} k_j(m_j^{(t)})}{\Omega} - \beta)$ is the penalty for violating the constraint condition where $\mathcal{C} \in \mathbb{R}^-$ is a constant. Finally, the reward function is the weighted combination of these two parts given by
\begin{equation}
    r^{(t)} = 
    wr_{g}^{(t)} + (1 - w)\sum_{i = 1}^{N}r_{i, l}^{(t)},
\end{equation}
where $w \in (0, 1]$ is the weight factor balancing the contribution of the utility from the problem objective function and the cost from the unsatisfied schedulability constraint. The values of $w$, $r_{pos}$, and $\mathcal{C}$ are hyperparameters and are tuned to produce the best learning outcome.
\end{itemize}
\subsubsection{Proposed DRL Framework}\label{DRL_proposed_framework}
In this part, the proposed centrally-trained-centrally-executed DRL based algorithm for solving the resource allocation problem is presented. The proposed multi-agent DRL based resource allocation algorithm is summarized in Algorithm \ref{alg:proposed_alg}, which is comprised of three parts, namely, initialization, random experience accumulation, and training and execution. 

\begin{algorithm}[!htb]
\caption{Proposed Multi-Agent DRL Based Resource Allocation Algorithm}\label{alg:proposed_alg}
\begin{algorithmic}[1]
    \Statex \textbf{Initialization:}
    \State Initialize the experience replay buffer, $\mathcal{D}$.
    \State For each node $i \in \{1,...,N\}$, randomly initialize the local DDQN network $\mathcal{Q}(s_i^{(t)}, a_i^{(t)}; \Psi_{i, agent}^{(t)})$, the train DDQN network $\mathcal{Q}(s_i^{(t)}, a_i^{(t)}; \theta_{i, agent}^{(t)})$, and the target DDQN network $\mathcal{Q}(s_i^{(t)}, a_i^{(t)}; \Bar{\theta}_{i, agent}^{(t)})$ with weights $\Psi_{i, agent}^{(t)}$, $\theta_{i, agent}^{(t)}$, and $\Bar{\theta}_{i, agent}^{(t)}$, respectively.
    \Statex \textbf{Random experience accumulation:}
    \State At the beginning of the time frame $t$, each node $i (\forall i \in \{1,...,N\})$ observes its local state  $s_i^{(t)}$ and transmits it to its corresponding network in the central server, and the server randomly chooses a blocklength $m_i^{(t)}$, and accumulates the local experience $e_i^{(t)}$ in the memory buffer $\mathcal{D}$. 
    \State Upon storing at least $\mathcal{B}$ experiences from different agents, a batch of size $\mathcal{B}$ is sampled from the replay buffer.
    \State The sampled mini-batch is used to compute the loss in (\ref{eq:DQN_loss}) and minimize it using an optimization technique.
    \State Update the train network parameters $\theta_{i, agent}^{(t)}$ using (\ref{gradient_update}).
    \State Update the target network parameters $\Bar{\theta}_{i, agent}^{(t)}$ using (\ref{soft_update}).
    \State Update DDQN parameters $\theta_{i, agent}^{(t)}$ to its corresponding local DDQN every time frame to update the weights $\Psi_{i, agent}^{(t)}$.
    \Statex \textbf{Training and execution:}
    \State Each node $i (\forall i \in \{1,...,N\})$ observes a local state $s_i^{(t)}$ and transmits the local observations to its corresponding network in the central core. The network executes the blocklength action $a_i^{(t)}$ following the $\epsilon$-greedy algorithm and stores the local experience $e_i^{(t)}$ in the replay buffer $\mathcal{D}$.
    \State A mini-batch of $\mathcal{B}$ experience is sampled from the buffer $\mathcal{D}$ to minimize the loss function in (\ref{eq:DQN_loss}).
    \State The train network parameters $\theta_{i, agent}^{(t)}$ are updated using (\ref{gradient_update}).
    \State The target network parameters $\Bar{\theta}_{i, agent}^{(t)}$ are updated using the soft update method in (\ref{soft_update}).
    \State Update DDQN parameters $\theta_{i, agent}^{(t)}$ to its corresponding local DDQN every time frame to update the weights $\Psi_{i, agent}^{(t)}$.
    \State Actions taken by the agents are sent back to each node to be executed. 
\end{algorithmic}
\label{alg1}
\end{algorithm}
In the initialization part, $N$ local DDQNs $\mathcal{Q}(s_i^{(t)}, a_i^{(t)}; \Psi_{i, agent}^{(t)})$ with network parameters $\Psi_{i, agent}^{(t)}$ are constructed based on the architecture of the proposed framework shown in Figure \ref{fig:arch}. Additionally, $N$ train DDQNs $\mathcal{Q}(s_i^{(t)}, a_i^{(t)}; \theta_{i, agent}^{(t)})$ and $N$ target DDQNs $\mathcal{Q}(s_i^{(t)}, a_i^{(t)}; \Bar{\theta}_{i, agent}^{(t)})$ with parameters $\theta_{i, agent}^{(t)}$ and $\Bar{\theta}_{i, agent}^{(t)}$ are, respectively, established. The weights of each network are initialized randomly, and the experience replay buffer $\mathcal{D}$ is initialized with its corresponding parameters (Lines 1-2).

In the random experience accumulation phase, at the beginning of time frame $t$, each node $i (\forall i \in \{1,...,N\})$ observes a local state $s_i^{(t)}$ and transmits it to its corresponding local network in the central server. Upon transmitting the information to the central server, each network randomly chooses a blocklength action $m_i^{(t)}$ to execute. The local network stores its local experience $e_i^{(t)}$ in the experience replay buffer $\mathcal{D}$ (Line 3). After collecting at least $\mathcal{B}$ experience samples from the agents, a mini-batch of size $\mathcal{B}$ is sampled and utilized to compute the loss value in (\ref{eq:DQN_loss}). The loss is minimized using the gradient update method in (\ref{gradient_update}) (Lines 4-5). After the gradient update, train network parameters $\theta_{i, agent}^{(t)}$ are updated in that time step, and target network is updated via a soft update method using (\ref{soft_update}) in that same time frame (Lines 6-7). Upon updating the training parameters in the central server, weights of the train network are sent to each local network in the central core every time frame to update the local network of each node (Line 8).

In the training and execution phase, agents execute their actions based on $\epsilon$-greedy policy rather than just choosing random actions. Each node observes a local state $s_i^{(t)}$ from the wireless environment and transmits it to the central server, and the network of each node executes a blocklength action and stores the local experience gathered from that interaction in the global replay memory buffer, and for the learning part, a mini-batch $\mathcal{B}$ is sampled to be processed to minimize the loss function (Lines 9-10). After minimizing the loss with the gradient update method used in (\ref{gradient_update}), train network parameters are adjusted, and the target network is updated using the same soft update method used in random experience accumulation (Lines 11-12). Finally, the parameters of the train network are utilized to update the parameters of the local agents, and the chosen actions in the execution phase are sent back to each node to be executed in the wireless environment (Lines 13-14).
\section{Performance Evaluation} \label{performanceevaluation}
 In this section, we evaluate the performance of the proposed optimization theory based DRL algorithm in comparison to the optimization based and pure DRL based benchmarks. In the optimization theory based benchmark, we develop an efficient approximation algorithm based on the further analysis of the optimality conditions, the relaxation of the resulting integer optimization problem and then searching of the integral solution by using a greedy algorithm, similar to the one proposed in \cite{sadi2014minimum} but adapted to the finite blocklength information theory. In the pure DRL approach, we adopt Branching Dueling Q-networks (BDQ) as the learning model to solve the optimization problem (13) without further analysis of the optimality conditions. There are three kinds of decision variables to optimize in order to have an optimal solution. To handle all three actions, continuous actions are discretized into $L$ levels. Applying DRL algorithms to high-dimensional action spaces increases the number of action dimensions. To overcome the problem of the complexity of commonly used sequential architectures, BDQ is proposed in \cite{tavakoli2018action} based on the usage of a shared decision model and several branching networks for each individual action dimension. The architecture of BDQ allows the selection of joint actions in a multidimensional setting without increasing the complexity of the network. 
 \vspace*{-\baselineskip}
 \subsection{Simulation Setup}
 Simulations are performed for a network where nodes are uniformly distributed within a circular area of a radius of $50$ meters, communicating with a central controller. The links between the sensor nodes and the central controller are impaired by both large-scale and small-scale fading. The large-scale fading $\alpha_i$ occurs due to path loss and shadowing and is modeled as $PL(d_i)$ = $PL(d_0) + 10 \alpha\log(\frac{d_i}{d_0}) + Z$ (in dB), where $d_i$ is the distance of node $i$ from the central controller, $PL(d_i)$ is the path loss of node $i$ at distance $d_i$ from the controller measured in decibels, $PL(d_0)$ = 35.3 dB is the path loss at reference distance $d_0 = 1 $ m, path loss exponent $\alpha=3.76$ \cite{access2010further}, $Z$ is a Gaussian random variable with zero mean and standard deviation equal to $4$ dB corresponding to log-normal shadowing. For the small-scale fading, Jake's model \cite{liang2017spectrum} is deployed, expressed as a first-order complex Gauss-Markov process:
 \begin{equation}
     f_{i}^{(t)} = \rho f_{i}^{(t-1)} + \sqrt{1 - \rho^2}e_{i}^{(t)},
 \end{equation}
 where $f_i^{(t)}$ and $e_i^{(t)}$ are channel coefficient and channel innovation process of node $i$ at time $t$, respectively; $\rho$ is the correlation coefficient. $f_{i}^{(0)}$ and the channel innovation process $e_{i}^{(1)}, e_{i}^{(2)}, ...$ are independent and identically distributed circularly symmetric complex Gaussian (CSCG) random variables with unit variance. $\rho$ is set to $0.6$. The channel gain of node $i$ at time frame $t$ is then computed by
 \begin{equation}
     g_{i}^{(t)} = |f_{i}^{(t)}|^2 \alpha_{i}, \quad t = 1, 2, ....
 \end{equation}
 The noise power spectral is $\sigma^2 = -174$ dBm/Hz. The parameters used in the simulations are given in Table \ref{table:parameters}.
\begin{table}[!htb]
    \centering
    \caption{Simulation parameters}
     \scalebox{0.9}{
         \begin{tabular}{|p{1.5cm}|p{1.5cm}|p{1.5cm}|p{2cm}|}
             \hline
             Parameter& Value &Parameter&Value\\
             \hline
             B   & $100$ kHz    &$M_{th}$& \footnotesize $200$ Symbols\\
             \hline
             L &   $100$ bits &  $\delta$ &  $0.99$\\
             \hline
             $\Delta$   & $1$ ms & $\Omega$ &$100$ ms \\
             \hline
              $W_{max}$ & $250$ mW & $W_c$  & $54$ mW \\
             \hline
              N & $50$ & $\sigma^2$  & \footnotesize $-174$ dBm/Hz \\
             \hline
             
        \end{tabular}
    }
    \label{table:parameters}
\end{table}
\vspace*{-\baselineskip}
\subsection{DRL Experimental Framework}
We next describe the neural network design and hyperparameters used for the architecture of our algorithm. Since our goal is to ensure that the agents make their decisions as quickly as possible, we do not over-parameterize the network architecture, and we use a relatively small network for training purposes. Our algorithm trains a DQN and DDQN with one input layer, three hidden layers, and one output layer. The hidden layers have $N_1 = 32$, $N_2 = 64$, and $N_3 = 300$ neurons, and the input layer has $3N + 3$ neurons. The leaky ReLU activation function is utilized to prevent the vanishing gradient problem. Furthermore, the activation function for the input and output layers is a linear function. In DQN, the number of neurons in the last layer is equal to the number of possible actions for every agent, denoted by $N_m$, which is equal to $M_{th}$. On the other hand, in DDQN, the final layer corresponds to the output of the advantage and state value functions. The number of neurons for the final layer is denoted by $N_A$ and $N_V$, which are equal to $M_{th}$ and $1$, respectively. A mini batch of size $32$ is sampled for the training phase for the all algorithms.

The pure DRL-based approach utilizes the BDQ algorithm to train and interact with the environment. In the output layer, the number of neurons dedicated to each output is chosen by the number of actions available for each output layer to take. The number of actions for the blocklength, sampling period, and packet error probability are $N_m$, $N_h$, and $N_p$, and are set to $200$, $512$, and $512$, respectively. The size of the input layer is $5N +3$. The Leaky ReLU is chosen as the activation function of the hidden layers, and a linear activation function is used for the input and output layers. The mini-batch size is set to $32$.

We use the RMSprop algorithm with an adaptive learning rate $\alpha^{(t)}$. For a more stable outcome, the learning rate is reduced as $\alpha^{(t+1)} = (1 - \lambda)\alpha^{(t)}$, where $\lambda \in (0, 1)$ is the decay rate of $\alpha^{(t)}$. Here, $\alpha^{(0)}$ is $0.03$ and $\lambda$ is $10^{-3}$. We also apply an adaptive $\epsilon$-greedy algorithm: $\epsilon^{(0)}$ is initialized to 1, and it follows $\epsilon^{(t+1)} = (1 - \beta)^{(t)} \epsilon^{(0)}$, where $\beta = 10^{-4}$. Discount factor ($\gamma$) in DRL setting is set to a moderate value of 0.666, since the connection between the actions taken by the agent and its forthcoming rewards tends to diminish due to fading, resulting in a lower correlation. Policy and target networks are updated via the soft update method, with $\tau=0.001$.

The proposed algorithm is implemented in PyTorch \cite{paszke2019pytorch}. Training is done by building the proposed architecture with the network parameters introduced in this section. Each testing result is an average of $10$ randomly initialized simulations of $2,500$ episodes.
\vspace*{-\baselineskip}
\subsection{Performance comparison and analysis}
Fig 3.a shows the training convergence of the reward for different algorithms in a network of 50 nodes with $\Omega = 0.1$ and $\Delta = 0.001$. The proposed and benchmark algorithms are shown to perform better than random selection, which selects random actions in each time frame. The conventional optimization technique has the best performance compared to cutting-edge DRL methods. However, in order to have optimal results, full CSI is needed to generate results, whereas DRL methods can use delayed or non-perfect CSI. DQN method outperforms BDQ in terms of getting higher rewards and converging faster in the training phase because the utilization of the optimization theory to reduce the constraints and decision variables has facilitated for the DRL model to learn the environment model. Due to its robust architecture and its ability to capture patterns in a stochastic environment, DDQN performs the best among the DRL algorithms in terms of converging and getting sub-optimal rewards.
\begin{figure}[!htb]
    \centering
    \subfigure[]{\includegraphics[width=.4\textwidth, trim={0 0 0 1.5cm}, clip]{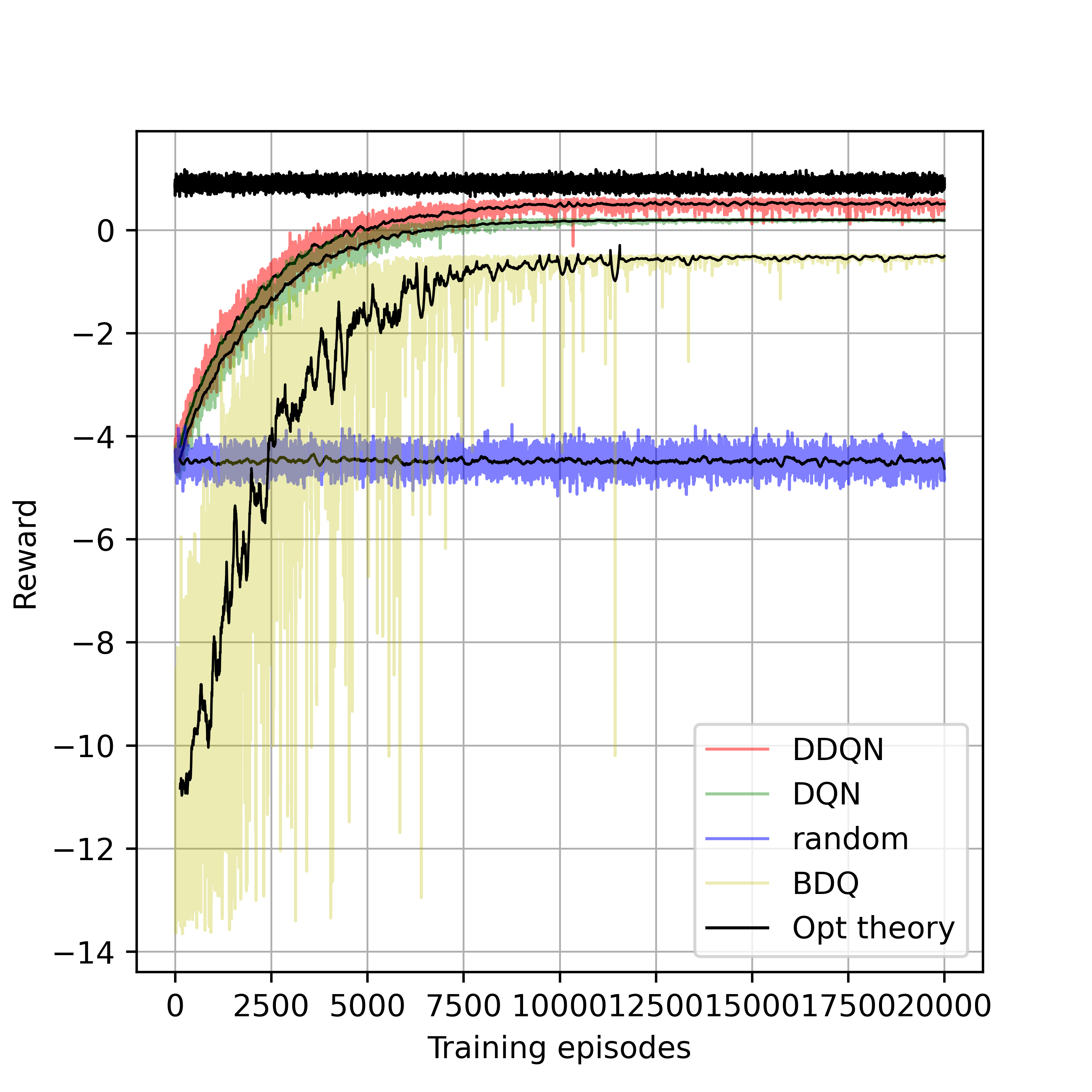}}\hspace{-0.1em}%
    \subfigure[]{\includegraphics[width=.4\textwidth, trim={0 0 0 1.5cm}, clip]{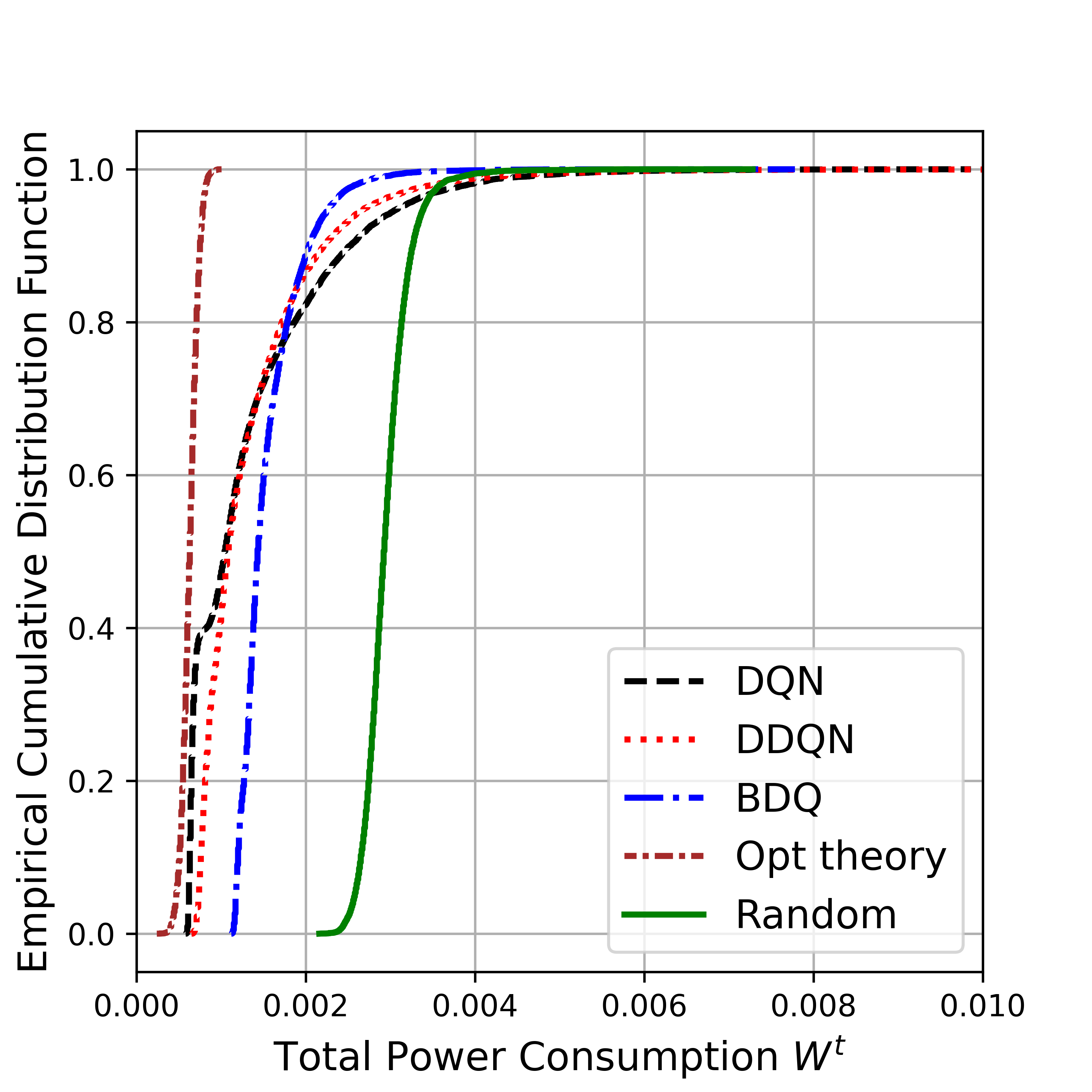}}
    \caption{Training and testing results for different algorithms in a network of $50$ nodes with $\Omega = 0.1$ and $\delta = 0.001$.}
    \label{default_results}
\end{figure}

 In Fig. 3.b, test results are shown for the proposed and benchmark algorithms. Testing is done based on the saved parameters of the trained networks in the training phase. The proposed optimization theory based DRL algorithms exhibit superior performance when compared to both benchmark approaches and random method. Nevertheless, they fall short of surpassing the capabilities of optimization theory. DDQN outperforms DQN in the testing part, which validates the training convergence graph. Similar to the training, both benchmark and proposed algorithms perform better than random selection. Similar to the training results, BDQ algorithm cannot outperform DDQN and DQN model due to its complexity and joint action selection feature.
\subsubsection{Effect of MATI}
In this part, we focus on the impact of varying the MATI value on the performance of the algorithms in the training and testing phase. Fig. 4.a shows the training convergence of the reward for different algorithms in a network of $50$ nodes with $\delta = 0.001$ and $\Omega = 0.06$. Similar to the previous results, all of the proposed and benchmark algorithms outperform the random-selection method. Similarly, DDQN outperforms DQN, which outperforms BDQ. The fluctuations of DQN and DDQN are less than that of BDQ, demonstrating their robustness. The BDQ algorithm fluctuates more than that in Fig. 3.a because of the more strict stochastic MATI constraint, which increases the cost function while training. Furthermore, the proposed algorithm performs very close to the pure optimization theory based approach. However, the performance of the proposed DDQN is within $25\%$ of that of the optimization theory based approach compared to $10\%$ in Fig. 3.a., due to the more strict stochastic MATI constraint.

Fig. 4.b shows the testing part of different algorithms. DDQN and DQN approaches significantly outperform random selection and BDQ algorithm. When the MATI condition is strict, there is no significant difference between BDQ and random selection method, which led to high fluctuations in the last episodes of the training phase. Lowering the MATI value leads to a rise in power consumption aimed at ensuring successful transmission under reduced MATI conditions.
\begin{figure}[!t]
    \centering
    \subfigure[]{\includegraphics[width=.4\textwidth, trim={0 0 0 1.5cm}, clip]{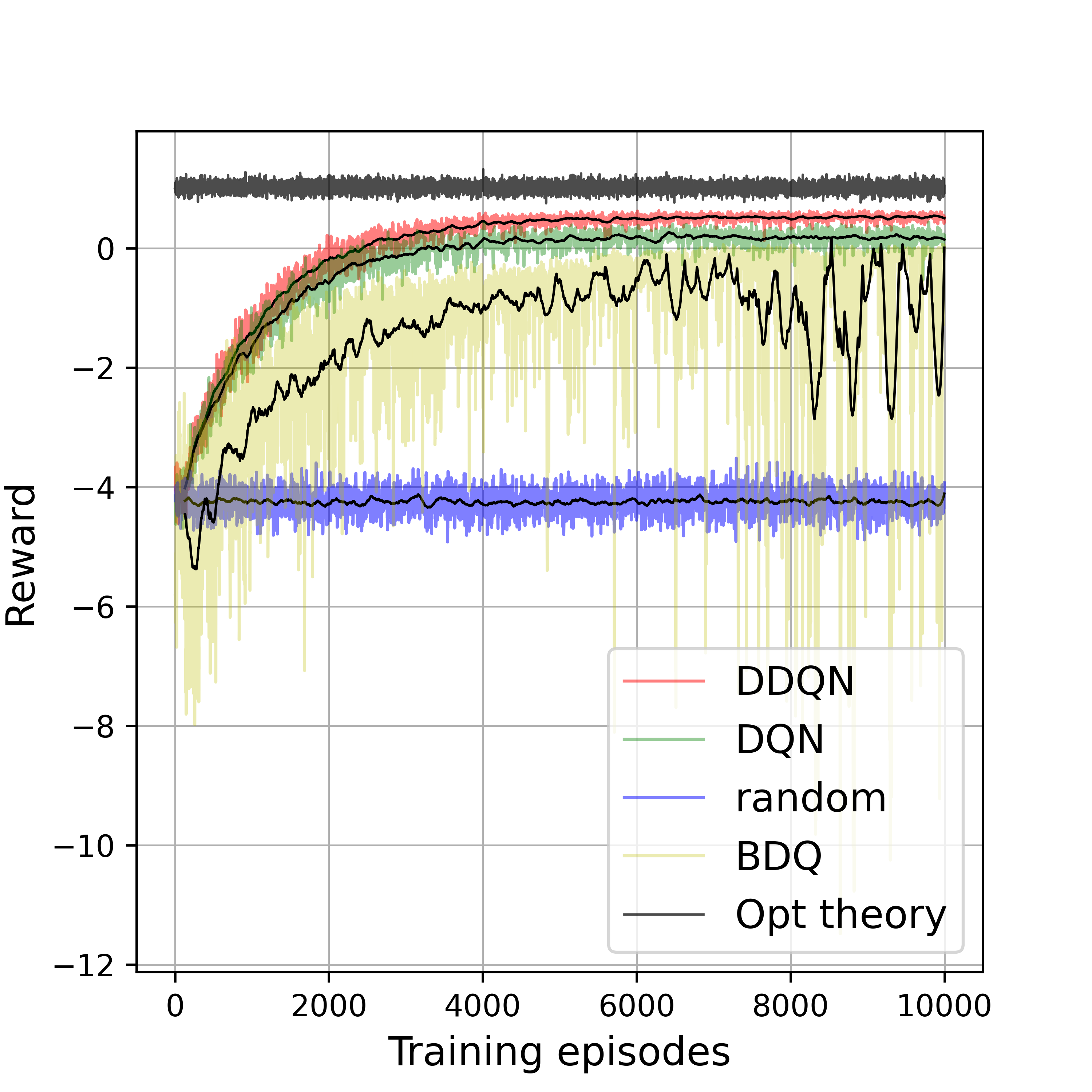}}
    \subfigure[]{\includegraphics[width=.4\textwidth, trim={0 0 0 1.5cm}, clip]{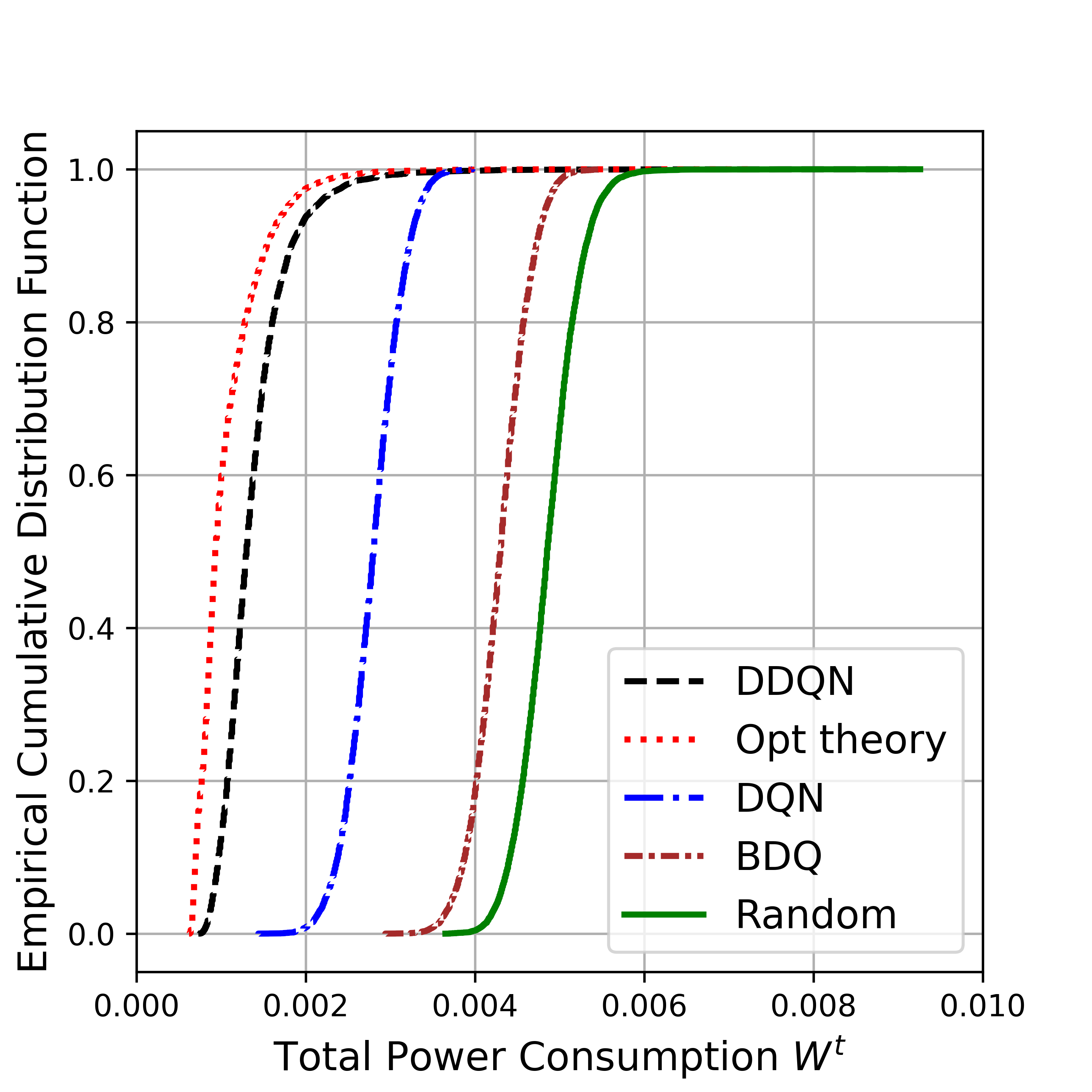}}
    \caption{Training and testing results for different algorithms in a network of $50$ nodes with $\Omega = 0.06$ and $\delta = 0.001$.}
    \label{MATI}
\end{figure}
\subsubsection{Effect of MAD}
Next, we analyze the impact of varying MAD value on the performance of the algorithms in the training and testing phases. Fig. 5.a shows the training convergence of the reward in a network of $50$ nodes with $\Omega = 0.1$ and $\delta = 0.01$. The increase in the MAD value relaxes the problem, which facilitates determining the optimal course of action. Consequently, the performance of BDQ is closer to that of DQN. On the other hand, DDQN outperforms DQN, and the disparity between them is more pronounced compared to prior scenarios. Moreover, the gap between the outcomes of optimization theory and the proposed DDQN method is more significant than in previous results (close to $50\%$ compared to $20\%$ of that in Fig. 3.a). This discrepancy arises from the fact that optimization results, assuming perfect CSI, yield suboptimal outcomes when the conditions within the system model are less stringent. However, DRL-based algorithms are less adept at efficiently capturing this effect than optimization methods when the conditions change slightly.  
\begin{figure}[!t]
    \centering
    \subfigure[]{\includegraphics[width=.4\textwidth, trim={0 0 0 1.5cm}, clip]{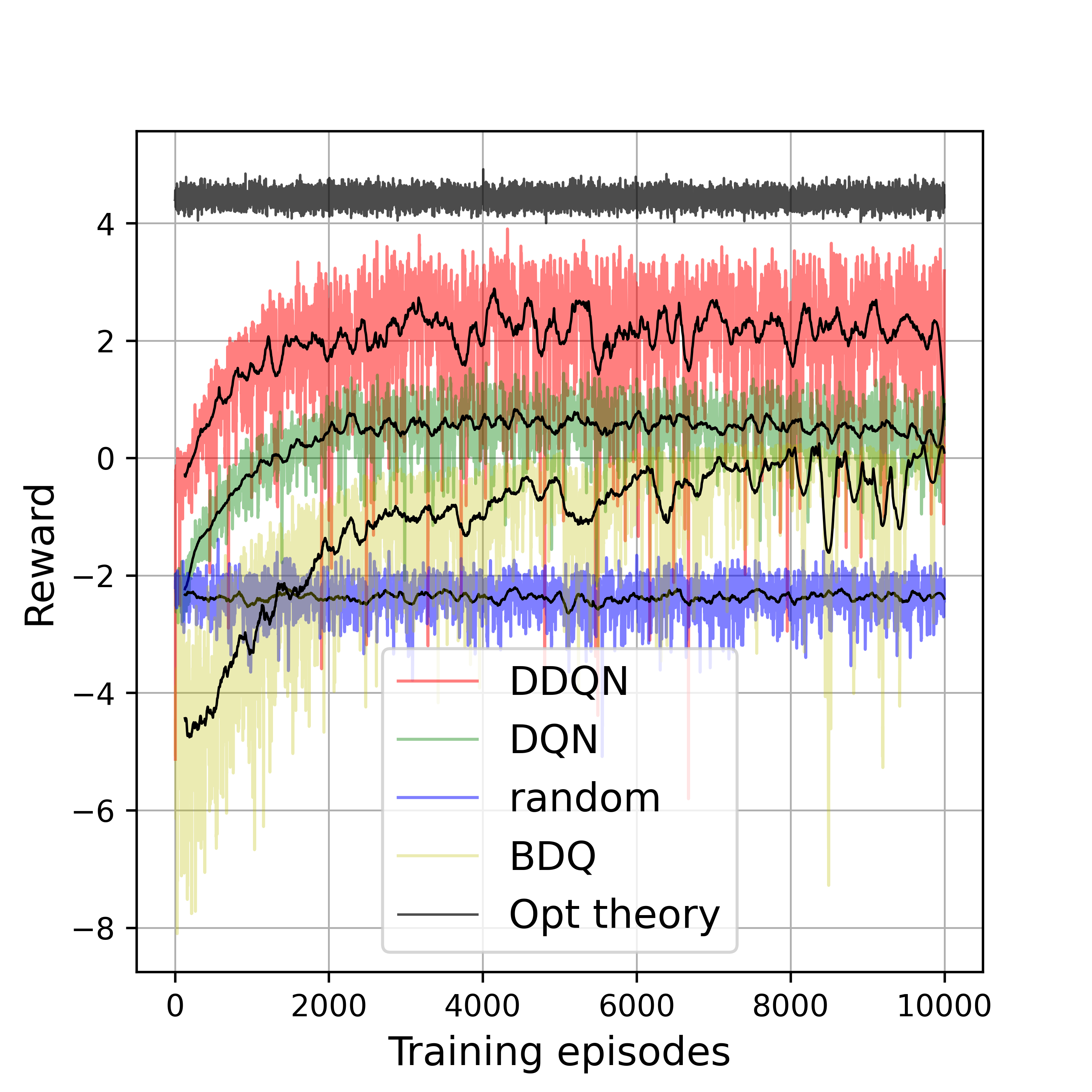}}
    \subfigure[]{\includegraphics[width=.4\textwidth, trim={0 0 0 1.5cm}, clip]{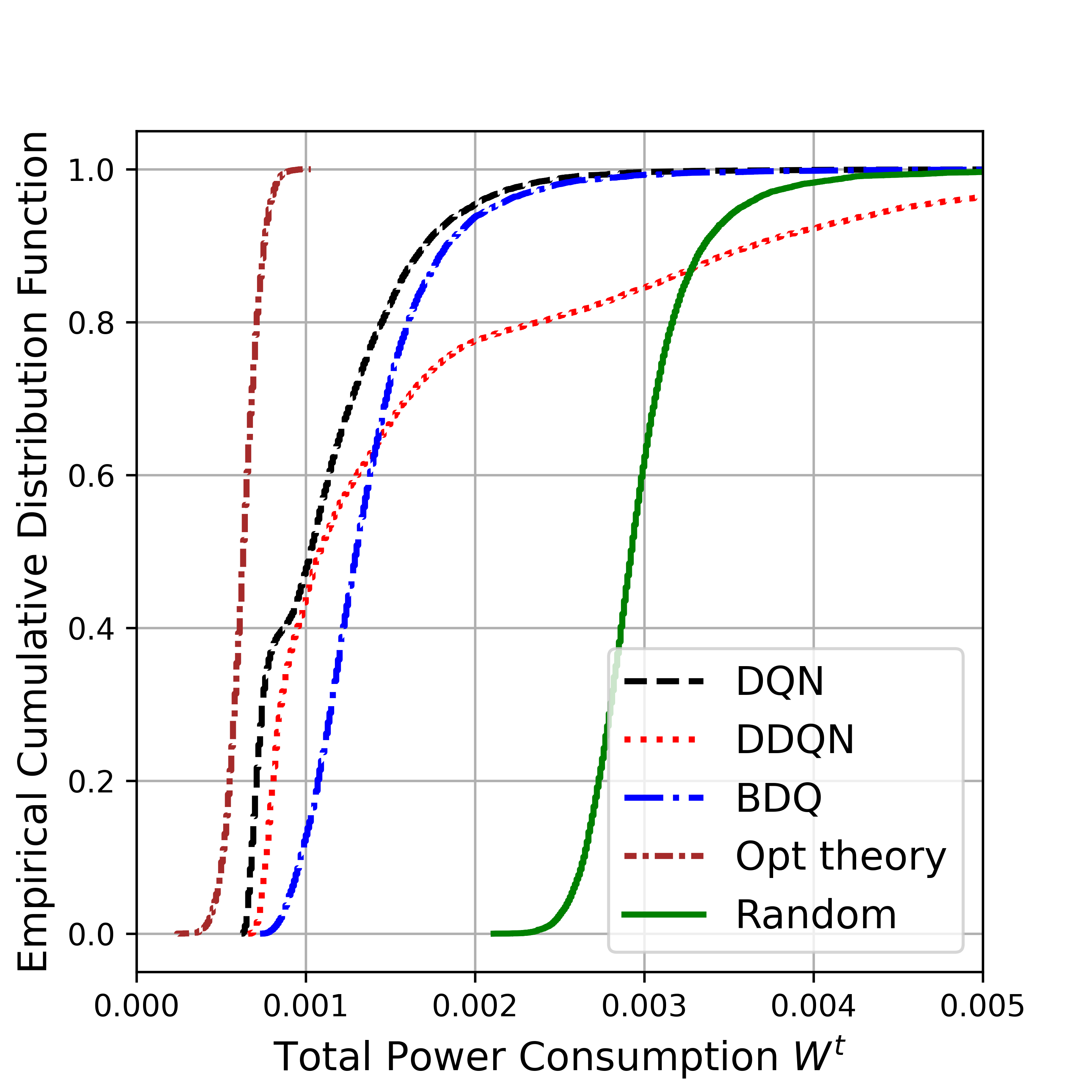}}
    \caption{Training and testing results for different algorithms in a network of $50$ nodes with $\Omega = 0.1$ and $\delta = 0.01$.}
    \label{MAD}
\end{figure}

Fig. 5.b shows the testing phase of the different algorithms with similar settings as the training part. Since the MAD constraint is more relaxed, the power consumption values are smaller, and there is no need to increase power in the control system. On one side, akin to previous situations, the suggested techniques demonstrate the capability to surpass random-selection and benchmark schemes. Conversely, BDQ algorithm is able to perform near to DDQN and DQN algorithms when compared to the previous results. As opposed to the previous results, DDQN results fluctuate more than DQN total power consumption results. The reason is that with a more relaxed scenario, an intricate approach like DDQN can overfit in the training phase, preventing the extraction of the best outcome in the testing part. 
\subsubsection{Effect of the Number of nodes}
We analyze the effect of the number of nodes on the performance of the proposed algorithms. Fig. 6.a shows the training convergence of the reward in a network of $25$ nodes, with $\Omega = 0.1$ and $\delta = 0.001$. Like previous setups, the proposed DDQN and DQN algorithms outperform the benchmark BDQ method. The optimization theory-based approach performs best since it has the perfect CSI of the communication model. The network complexity is reduced compared to the scenario with $N = 50$, resulting in minor differences between the algorithms compared to the previous case. During the training phase, random selection proves to be a subpar strategy due to its inadequate ability to select optimal power values. Conversely, the optimization method excels when complete CSI is available. DRL methods showcase performance near optimality without requiring full CSI. The proposed DDQN method is able to perform about $20\%$ close to the optimization method when the network size is decreased. 
\begin{figure}[!htb]
    \centering
    \subfigure[]{\includegraphics[width=.4\textwidth, trim={0 0 0 1.5cm}, clip]{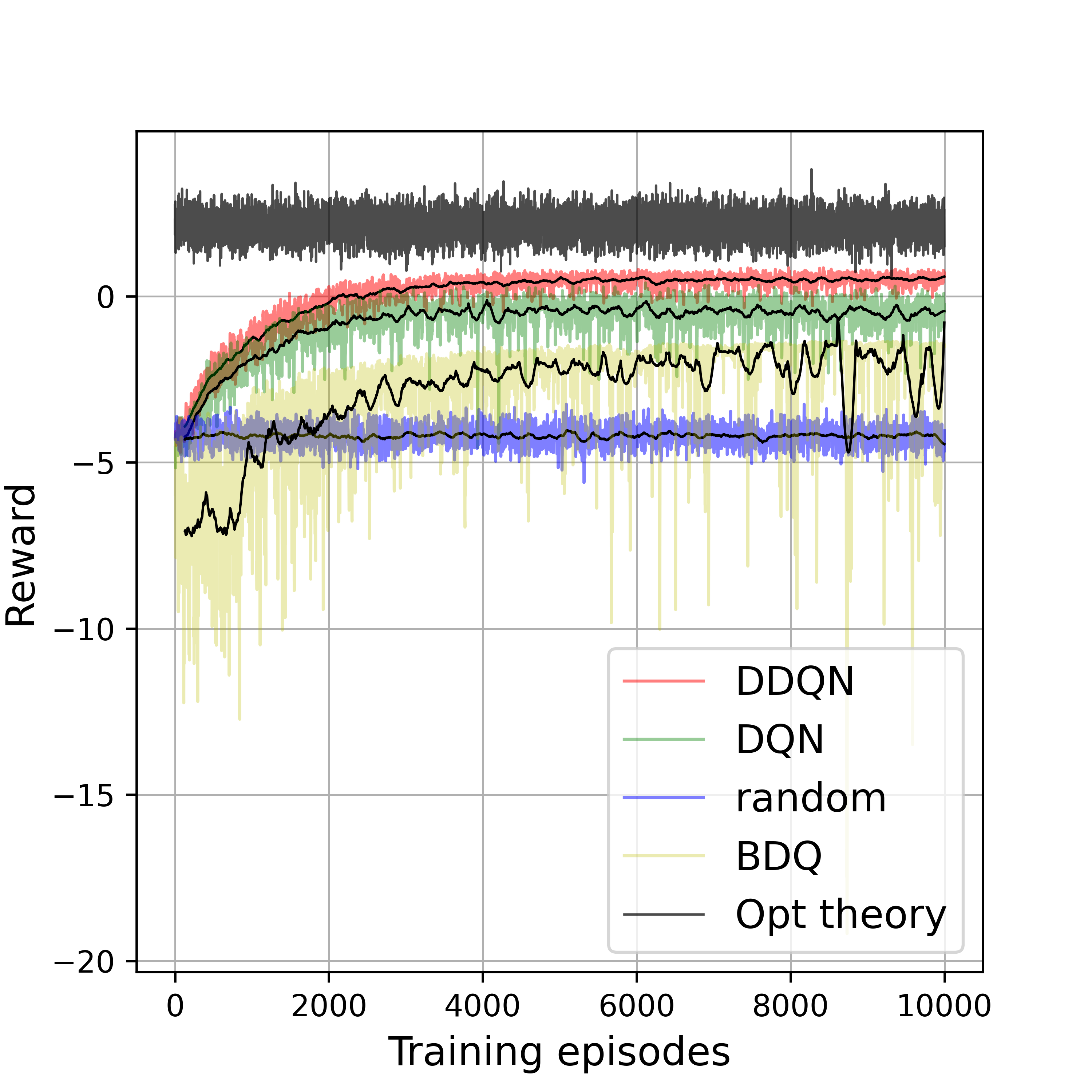}}
    \subfigure[]{\includegraphics[width=.4\textwidth, trim={0 0 0 1.5cm}, clip]{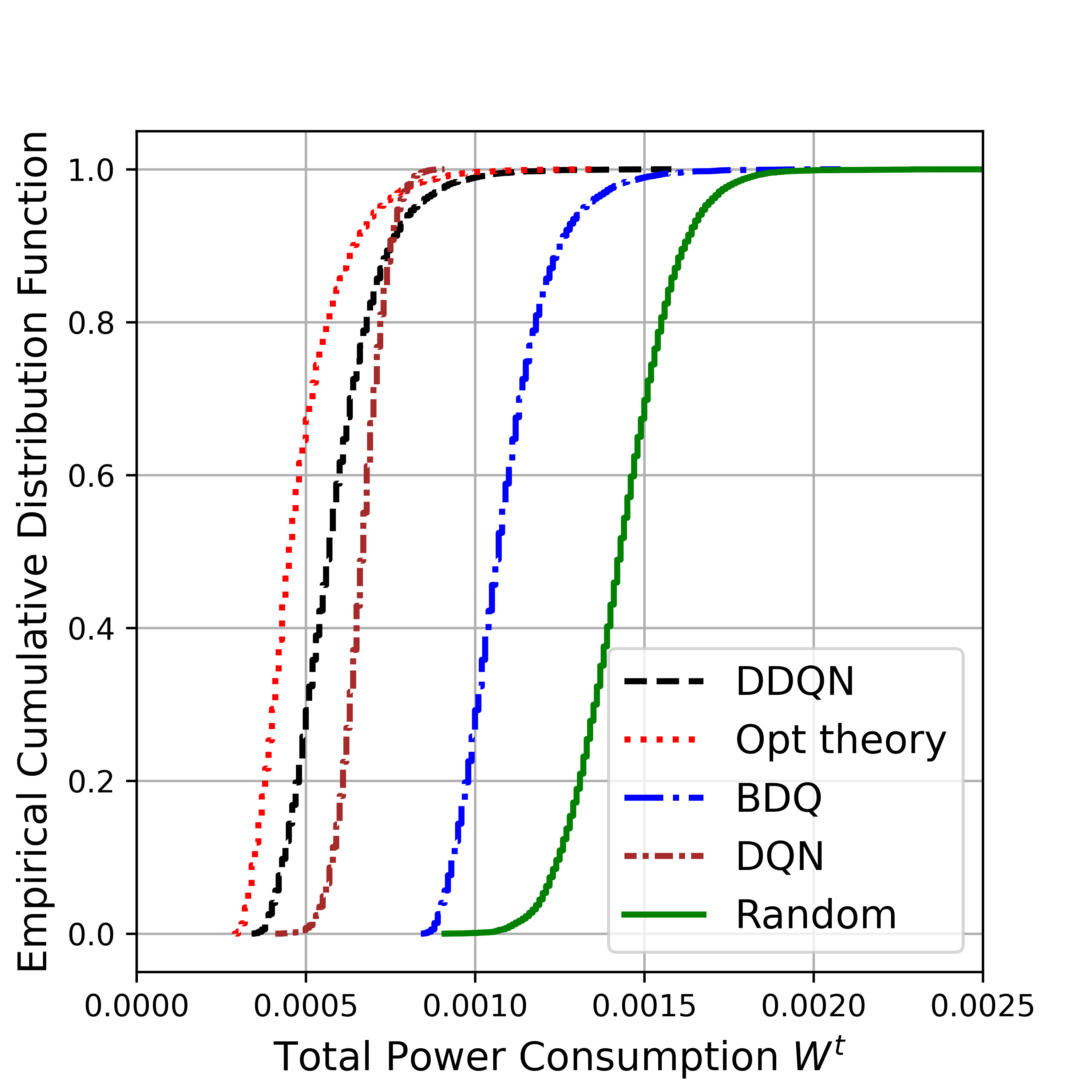}}
    \caption{Training and testing results for different algorithms in a network of $25$ nodes with $\Omega = 0.1$ and $\delta = 0.001$.}
    \label{Nodes}
\end{figure} 
Fig. 6.b shows the testing phase of the network where DDQN and DQN perform close to the optimization theory based solution because of the decreased complexity of the system. The power consumption decreases as $N$ decreases. Among the DRL algorithms, DDQN and DQN exhibit the best performance, and much better than BDQ and random selection.

\begin{figure}[!htb]
    \centering
    \includegraphics[width=.4\textwidth, trim={0 0 0 1.5cm}, clip]{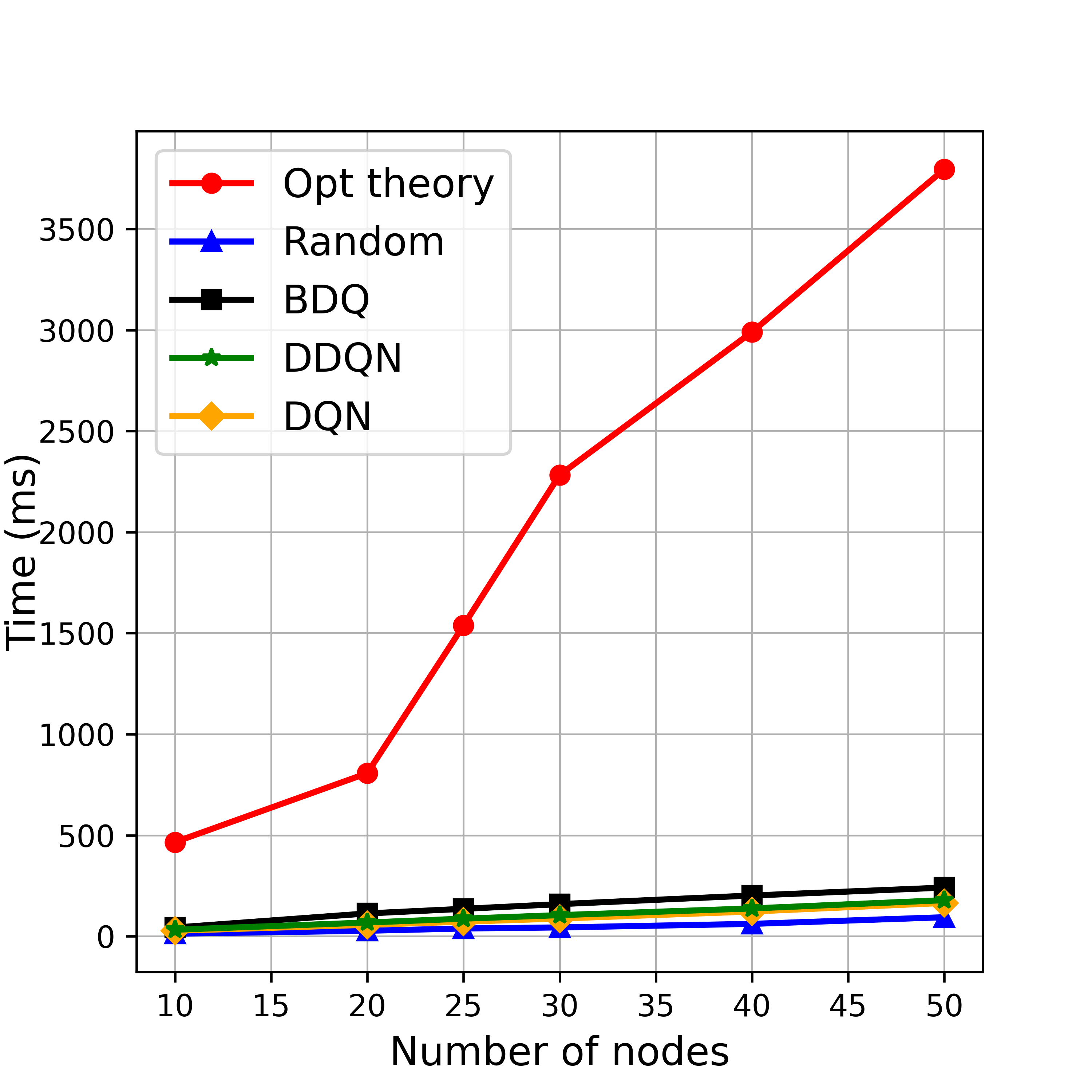}
    \caption{Average execution time (ms) per step for different algorithms and number of nodes.}
    \label{fig:time}
\end{figure}

Fig. \ref{fig:time} shows the average execution time for each iteration of the algorithms. A substantial disparity exists between the optimization method and DRL-based approaches as the number of nodes increases. This disparity indicates that DRL-based methods can be effectively harnessed for real-time applications with close to optimal results.  

\section{Conclusion} \label{conclusion}
In this paper, we propose a novel optimization theory based DRL framework for the joint optimization of control and communication systems with the goal of minimizing power consumption while considering ultra-reliable transmission in the finite blocklength regime. Following the formulation of the optimization problem based on the abstraction of the control and communication systems, the proposed methodology derives the optimality conditions to decompose the problem into multiple building blocks. Then, the building blocks that are not tractable are replaced by DRL, which employs DQN and DDQN to choose optimal actions. The proposed optimization theory based DRL algorithms, DQN and DDQN, have been demonstrated to outperform the benchmark pure DRL-based algorithm, BDQ, and the random selection method while performing very close to the conventional optimization theory based approach for different network sizes, MATI, and MAD values. In the future, we plan to incorporate alternative control system abstractions, such as Age-of-Information, Age-of-Loop, into the proposed optimization theory based deep learning solution methodology and analyze the interpretability and robustness of the proposed algorithms through the development of explainable AI techniques.

\bibliographystyle{ieeetr}
\bibliography{main}

\end{document}